\documentclass[a4paper,11pt]{article}

\usepackage{aaskaiid}
\usepackage{orcidlink}

\include{journal-names}

 \usepackage{natbib}

\title{Searching for Extraterrestrial Intelligence with the SKA }
\ShortTitle{Technosignatures with the SKA}
\author[1,2]{Chenoa D. Tremblay\orcidlink{0000-0002-4409-3515}}
\ShortName{Tremblay et al.} % shortened name list for header 
\emailAdd{ctremblay@seti.org}
\author[5]{Alex Andersson\orcidlink{0000-0003-2734-1895}}
\emailAdd{alexander.andersson@physics.ox.ac.uk}
\author[5]{Joe Bright\orcidlink{0000-0002-7735-5796}}
\emailAdd{joe.bright@physics.ox.ac.uk}
\author[6]{B\'arbara Cabrales\orcidlink{0009-0009-1019-3890}}
\emailAdd{bcabrales@seti.org}
\author[1,3,5]{David DeBoer\orcidlink{0000-0003-3197-2294}}
\emailAdd{ddeboer@berkeley.edu}
\author[1]{Vishal Gajjar\orcidlink{0000-0002-8604-106X}}
\emailAdd{vishalg@berkeley.edu}
\author[6]{Michael A. Garrett\orcidlink{0000-0001-6714-9043}}
\emailAdd{michael.garrett@manchester.ac.uk}
\author[4]{Evan F. Keane\orcidlink{0000-0002-4553-655X}}
\emailAdd{evan.keane@tcd.ie}
\author[7]{Dong-Jin Kim}
\emailAdd{Dongjin.Kim@csiro.au}
\author[5]{Steve Prabu}
\emailAdd{steve.prabu@physics.ox.ac.uk}
\author[8]{Danny C. Price\orcidlink{0000-0003-2783-1608}}
\emailAdd{Daniel.Price@skao.int}
\author[1,5]{Andrew P.V. Siemion\orcidlink{0000-0003-2828-7720}}
\emailAdd{andrew.siemion@physics.ox.ac.uk}
\author[1,3]{Sofia Z. Sheikh\orcidlink{0000-0001-7057-4999}}
\emailAdd{ssheikh@seti.org}
\author[13,14]{Cyril Tasse\orcidlink{0009-0009-9030-7885}}
\emailAdd{cyril.tasse@obspm.fr}
\author[12,13]{Philippe Zarka\orcidlink{0000-0003-1672-9878}}
\emailAdd{philippe.zarka@obspm.fr}
\author[1]{Nathalie Cabrol\orcidlink{0000-0001-7520-4364}}
\emailAdd{ncabrol@seti.org}
\author[10,11]{Joseph R. Callingham\orcidlink{0000-0002-7167-1819}}
\emailAdd{callingham@astron.nl}
\author[1,3,5]{Steve Croft\orcidlink{0000-0003-4823-129X}}
\emailAdd{scroft@berkeley.edu}
\author[1]{William Diamond}
\emailAdd{bdiamond@seti.org}
\author[5]{Jamie Drew}
\emailAdd{drew@breakthroughprize.org}
\author[9,5]{Kyran Grattan}
\emailAdd{grattan@breakthroughprize.org}
\author[9]{S.Peter Worden}
\emailAdd{pete@breakthroughprize.org}
\affiliation[1]{SETI Institute, 339 Bernardo Ave, Suite 200, Mountain View, CA 94043, USA}
\affiliation[2]{Department of Physics and Astronomy, University of New Mexico, Albuquerque, NM 87131, USA}
\affiliation[3]{Berkeley SETI Research Center, Berkeley, CA 94720, USA}
\affiliation[4]{School of Physics, Trinity College Dublin, College Green, Dublin 2, D02 PN40, Ireland}
\affiliation[5]{Sub-department of Astrophysics, University of Oxford, Oxford, OX1-3RH, UK}
\affiliation[6]{Jodrell Bank Centre for Astrophysics, Dept. of Physics \& Astronomy, Alan Turing Building, Oxford Road,
University of Manchester, M13 9PL, UK.}
\affiliation[7]{Australia Telescope National Facility, CSIRO, Space and Astronomy, PO Box 76, Epping, NSW 1710, Australia}
\affiliation[8]{SKA Observatory, Science Operations Centre, Kensington, WA 6151, Australia}
\affiliation[9]{Breakthrough Initiatives, Steward Observatory, The University of Arizona, Tucson AZ 95719, USA.}
\affiliation[10]{ASTRON, Netherlands Institute for Radio Astronomy, Oude Hoogeveensedijk 4, Dwingeloo, 7991\,PD, The Netherlands}
\affiliation[11]{Anton Pannekoek Institute for Astronomy, University of Amsterda, Science Park 904, 1098\,XH, Amsterdam, The Netherlands} \affiliation[12]{LIRA, Observatoire de Paris, Universit\'e PSL, CNRS, 92190 Meudon, France}
\affiliation[13]{Observatoire Radioastronomique de Nan\c cay, Observatoire de Paris, CNRS, PSL, UO, 18330 Nan\c cay, France}
\affiliation[14]{LUX, Observatoire de Paris, Universit\'e PSL, CNRS, 92190 Meudon, France}
\begin{document}
\maketitle

\begin{abstract}
    TThe search for technosignatures (also known as the Search for Extraterrestrial Intelligence or SETI) depends critically on our ability to distinguish artificial signals from the rich complexity of natural astrophysical phenomena and radio frequency interference from anthropogenic emissions. As the search for technosignatures increasingly aligns with mainstream astrophysics, complementing the search for biosignatures, it demands not only sophisticated statistical and computational approaches, but also deep domain knowledge across the electromagnetic spectrum. The SKA will play a pivotal role in the next-generation of technosignature searches, providing an unprecedented combination of sensitivity, field of view, and spatial resolution over its wavelength range.
Integrating wide-field, high-resolution observations with machine learning and multi-wavelength diagnostics will represent key steps forward. The SKA’s singular capabilities will render it an indispensable instrument for the rapid identification and follow-up characterisation of promising technosignature candidates. In this chapter, we discuss the request for high temporal and spectral resolution data products with a main focus on frequency-domain SETI. Nevertheless, multiple SKA observing modes have the potential to substantially advance technosignature research.
\end{abstract}
%\tableofcontents
\clearpage

\section{Introduction}
One of the primary goals stated by the SKA Cradle of Life working group\footnote{\url{https://www.skao.int/en/science-users/science-working-groups/105/cradle-life}} is to advance the quest to find the proof of life; biological or technological. This is a central scientific and philosophical issue that has persisted throughout human history, and the intensity of the search for an answer will not diminish at any time in the foreseeable future. At the moment, we only know one planet that definitively hosts life: Earth. However, with recent investigations into potential Hycean worlds (e.g., \citealt{Barclay_2021,Madhusudhan_2023}), the potential for phosphine on Venus \citep{Mrazikov_2024}, the discovery of amines on asteroids \citep{Qasim_2023}, and the chemical complexity of protoplanetary disks (e.g., \citealt{Potapov_2025}), the field of astrobiology that underscores the search for life is more optimistic than ever about finding evidence of life. Knowing that life could exist is different from discovering that life does exist. \citet{Wright_2022} argue that we can bypass theoretical debates about the definition of life and what it takes for it to exist, by searching for signs of technology designed and functioning in another part of the cosmos (technosignatures).  This philosophy echoes sentiments voiced in the early days of the modern search for technosignatures, in which Carl Sagan et al. stated that ``we are unanimous in our conviction that the only significant test of the existence of extraterrestrial intelligence is an experimental one. No a priori arguments on this subject can be compelling or should be used as a substitute for an observational programme.'' 

The quest to find intelligent or technologically capable life in the Universe has experienced significant growth in the last decade due to funding, renewed government agency interest, and emerging technologies. In particular, since the last SKA Science chapters were published~\citep{SKA_Book}, more than 200 articles and conference abstracts have been published on technosignature research, with the rate increasing each year. In this chapter we focus on these works. These, which span instrumentation and experimental design (e.g. \citealt{Huang_2023, Price_2024}), theory (e.g. \citealt{Hippke_2017,Haliki_2019,Gray_2020,Li_2022,Li_2023,Kipping_2025}), survey strategies (e.g. \citealt{Houston_2021,Tusay2024}), and numerous observational surveys (e.g.~\citealt{Johnson_2023}). The surveys have even expanded beyond the traditional stellar-focused narrowband signal searches (e.g., \citealt{Isaacson_2017,Enriquez_2019TT,Price_2020,BLC1,Tao_2022,Painter_2024}) to include extragalactic sources \citep{Garrett_2023,Choza_2023,Uno_2023,Tremblay_EG}, pulsed signals \citep{Perez_2020,Gajjar_2021,Suresh_2023}, scintillating signals \citep{Brzycki_2023,Brzycki_20204}, and broadband signals (e.g. \citealt{Tremblay_2022_SETI,Gajjar_2022}). The published literature also now contains historical information on international SETI efforts \citep{Gindilis_2019}, illuminating past surveys and the foundations of social science in the theory and history of SETI. Although there has been a great deal of growth in this field, the greatest challenge remains to expand the sensitivity, sky coverage, emission modalities, and signal detection and filtering algorithms to set more stringent limits on the prevalence of technological life in the Universe. For this, SKA will be an invaluable resource.

Radio astronomy offers unique advantages in the search for technosignatures. Radio emissions are cost-effective per photon, and radio waves pass through the intergalactic, interstellar, and interplanetary mediums relatively undistorted and unimpeded, apart from plasma effects \citep{Cordes_1991,Siemion_2013,Tremblay_2022_SETI}. Many papers have discussed the benefits of searching for life with radio telescopes (e.g., \citealt{Oliver_1971,Tarter_2003,Shuch_2011}). The most widely discussed advantages include the following: 
\begin{itemize}
    \item coherent radio emission is commonly used (on Earth) to facilitate our furthest and most complex communication signals;
    \item electromagnetic transmissions travel at the speed of light, representing our fastest method of communication across significant distances;
    \item radio photons are low energy and economical (energetically) to produce;
    \item some sources of coherent radio emissions can be distinguished from sources of astrophysical background with some ease;
    \item extraterrestrial leakage and beacons, as received, would be unlikely to be constant in frequency due to relative Doppler motion, providing an additional discriminant in any detection. 
\end{itemize}
All of these features are agnostic to the purpose or motivation of the emissions that could be received. For example, it does not matter whether the signal was intentional or unintentional in nature, and the search is relevant across a variety of different types and modulations of signals. 

Current values suggest that in certain parts of the radio spectrum the Earth's emission outshines the Sun by a factor of more than a million \citep{SETI2020}. Thus, we may expect that the leakage from other civilisations is unlikely to be a single coherent narrow-band emission \citep{Sullivan_1978,Saide_2023,Haqq_2025,Sheikh_2025}, but the collective sum of many radio transmissions. \cite{Saide_2023} explored this from the perspective of detecting mobile phone networks from our nearest stars. The authors found that the signals generated by the mobile towers are variable due to their non-uniform distribution around the Earth but that our stellar neighbour HD\,95735 could detect us at a level of up to 4\,GW. In other words, using the calculations from \citep{Sheikh_2025}, an SKA-Mid-like telescope up to 4.0~light years from Earth would be able to detect our LTE cell emissions with only an hour of integration time. 

\cite{Haqq_2024} take a more inclusive approach by studying the Earth's ``technosphere" as a whole and the properties it may have when observed from afar. They also provide estimates on the expected emission power of the technosphere up to 1000 years in the future through a series of different scenarios. They estimate that currently we are emitting about $6 \times 10^{20}$\,J of radiation continuously, but suggest that this value will probably decrease over time due to the increasing thermal efficiency of our electronics. However, as we progress toward having radio telescopes on the Moon (e.g., \citealt{Lunar_VLBI}), we can get an even better understanding of the true values for the Earth's technospheres \citep{Hibbard_2025} to develop more effective search strategies to find Earth-like technology on other planets by looking down at Earth from afar.\footnote{We currently have some sense from RFI reflected off the moon, but this is a limited view from strong signals. A view from the moon looking at Earth would give a more complete picture of the technosphere.} 

The primary limitation to building a strong radio emitter would probably be the energetic resources to power it \citep{Sheikh_2025}. This suggests that our best chance of detection is to search for either weak but continuous signals or very strong signals from intermittent sources of emission, instead of an extremely bright (and therefore expensive) continuous emitter. To explore these weak or intermittent regimes, our searches need more sky coverage, better sensitivity, wider bandwidths, and overall a lower cost per volume of parameter space: all of which are offered by the SKA.  

The theoretical motivation discussed by \cite{Cocconi_1959} and Frank Drake's experiments conducted on the 85-foot telescope in Green Bank, West Virginia~\citep{drake_1961}, are considered to be the beginning of observational SETI.
In those early days, surveys were focused on narrowband, drifting signals around the spectral region known as the `water hole' (1.2--1.7\,GHz; the frequency band between H{\sc I} and OH emission). In SETI, `narrowband' is defined as a signal with a bandwidth smaller than that produced by any known astrophysical phenomena (e.g., masers with $\sim$500\,Hz spectral width). The narrowest possible signal would be a pure continuous-wave carrier, carrying no encoded information; this kind of signal would essentially contain a single bit of information: technology is present.

Today, our search strategies and frequency bands expand to the limits of our imagination and understanding of electromagnetic radiation propagation. Ideally, we would like to find rich messages full of information about civilisation. For these styles of communication, broadband ($\sim$10s of kHz) signals would be required (e.g. \citealt{Clancy_1980, Gajjar_2022}). Another consideration to ensure detection would be to employ a modulation mode that contains individual narrow-band components scattered across a wider stretch of the electromagnetic spectrum \citep{Suresh_2023}. This would assume that the other civilisations were purposefully trying to signal and draw attention to themselves. More passive escaping signals from a civilisation's day-to-day life would be broadband generated, potentially as they communicate among themselves.

We do not know the frequency at which a signal would be emitted by other technologically capable civilisations. For example, \cite{Kardeshev_1979} considered the search for call signs (beacons) and informative broadcasts and suggested that the optimum frequency for directed radiation was at 1.5\,mm (203\,GHz) near the positronium line.  Many of the older SETI experiments were based on what was readily available to the observer constrained by the development of electronics --- not necessarily about which form of communication was most efficient. \cite{Hippke_2017} discuss this exact problem. They suggested that the main concern presented by groups such as Project Cyclops \citep{Cyclops}, a group in the United States assembled to discuss search strategies, was that the signal would be indistinguishable from black-body radiation. However, astrophysical objects are not perfect blackbody emitters, suggesting that if the technosignature search is completed at high resolution, an astrophysical object should have characteristic absorption lines from elements and molecules, where technosignatures would not. The experiments for technosignatures over the last decade have branched out to cover a significant portion of the radio wavelength range observable by ground-based telescopes. This has included ranges of 10 to 190 MHz (e.g., \citealt{Tingay_2016,nenufar,Johnson_2023}), 500\,MHz to 2\,GHz (e.g., \citealt{Bowyer_1995,Czech_2021,Sheikh_2021,Margot_2023}), 2 to 50\,GHz (e.g., \citealt{Slysh_1985,Gajjar_2022,Sheikh_2020,Manunza_2024}), and 90.642 to 93.151~GHz \citep{Mason_2025}\footnote{A list of searches and their covered frequencies are compiled by the community at \url{https://technosearch.seti.org/radio-list}}. 

Note that there is no physical limitation on the effective power of a beamed phased array in the electromagnetic spectrum. 
However, microwave transmitters can outshine the Sun, using powers of megawatts and narrow bandwidths, making this a compelling experiment for the next generation technology. 

\begin{figure}
\includegraphics[width=1.0\textwidth]{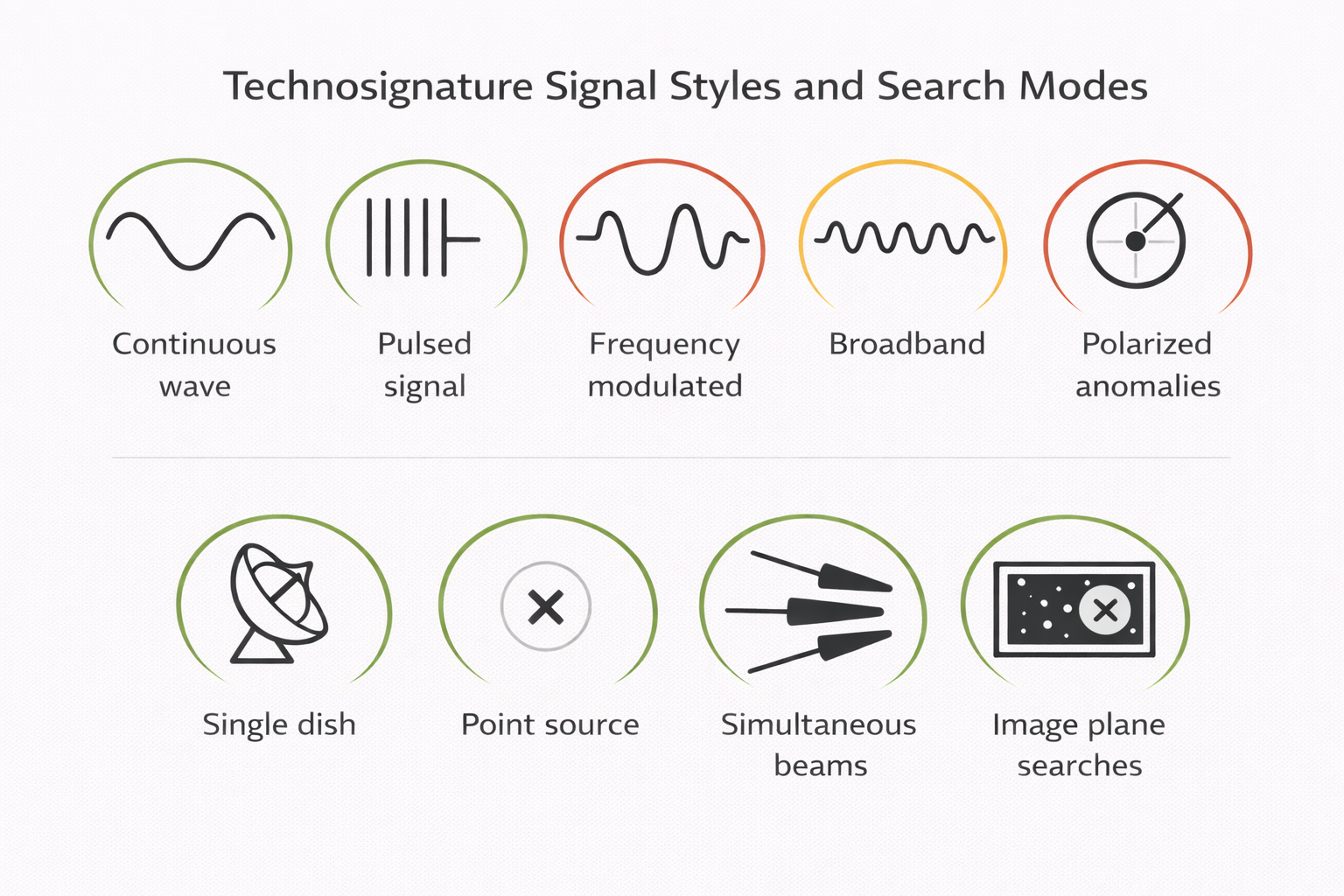}
\caption{With the SKA we can open up a wide range of sensitive searches for technosignatures of various types and styles. Although most searches involve looking continuos narrowband signals, as we develop the SETI field and SKA capabilities come online, we can expand the search to cover more parameter space. The green circles represent active areas of research and are discussed in this chapter, the yellow are areas of research currently under investigation, and the red circles are search types we hope to explore as more SKA capabilities come online. \label{fig:summary}}
\end{figure}

In this chapter, we explore the search for technosignatures—artificial signals that can indicate the presence of technological activity beyond Earth. We outline the range of signal morphologies examined in contemporary technosignature research and highlight the unique capabilities the SKA offers to advance this field. In Figure \ref{fig:summary}, we demonstrate the increased capabilities in the field due to the addition of the SKA's capabilities.

\section{Technosignature Observations with SKA}
The SKA combines unprecedented sensitivity, a broad frequency range, and a wide field of view, offering capabilities that surpass those of current radio telescopes. These features dramatically enhance our ability to detect potential technological signatures beyond Earth. In this section, we summarise preparatory work carried out with SKA precursor and pathfinder facilities and discuss how the full SKA will extend and refine these efforts.

\subsection{Observations with SKA Low}
If we assume a signal that is drifting in frequency due to radial planetary motions (i.e., orbital and rotational motions), then the total noise (proportional to the bandwidth) increases by a factor of square root of the frequency simply because of the expected Doppler drifts \citep{Sheikh_2019,Li_2023}. This favours the low end of the radio frequency band, as lower frequencies will experience less noise from this effect. Additionally, because of the large size of the wave, the emission is less likely to be blocked by planetary atmospheres, the interstellar medium, or dust as it travels to our transmitters on Earth \citep{Tremblay_2022_SETI}. However, the host star’s exoplanetary interplanetary medium (Exo–IPM) is likely to induce spectral broadening on the order of $\sim$10–100\,Hz for narrowband signals, an effect that can be mitigated by employing width-aware narrowband search algorithms \citep{2025AAS...24611707B}. 

Low-frequency radio telescopes possess several unique characteristics compared to previous SETI experiments conducted at higher frequencies, owing to their wide fields of view and low observing bands. As discussed by \citet{Garrett_2017}, both the Murchison Widefield Array (MWA; \citealt{Tingay_2013}) and the LOw Frequency ARray (LOFAR; \citealt{vanHaarlem_2013}) are aperture array instruments capable of surveying large regions of the sky—up to 900,deg$^2$ for the MWA and approximately 10,deg$^2$ for LOFAR—enabling simultaneous observations of millions of stars. 
This is also the case of the SKA pathfinder NenuFAR in Nan\c{c}ay \citep{Zarka2020}.
However, their synthesised beams are relatively large, limiting the ability to distinguish individual stellar systems within a single observation.

The low operational frequencies of these facilities — 80–300~MHz for the MWA, 10–80~MHz and 120–240~MHz for LOFAR, 10–85~MHz for NenuFAR and 50–350~MHz for SKA-Low — coincide with bands commonly used for terrestrial AM and FM radio, aircraft communication, high-power military satellites, and digital television transmissions \citep{Tremblay_2022_SETI}. The SKA-Low site in Western Australia benefits from an established radio-quiet zone \citep{OffriingaRFI,Sokolowski_17}, which, combined with rigorous RFI excision and data-quality checks, enhances the potential for detecting both intentional and unintentional radio emissions from exoplanetary systems that may host technology comparable to our own.

As precursors and pathfinder instruments, the MWA, LOFAR and NenuFAR
have demonstrated the feasibility of using the SKA-Low telescope for technosignature surveys. The MWA covered two aspects of life detection science simultaneously using correlated images: the search for biomolecule tracers and the search for intentional or unintentional technological signals \citep{Tremblay_PhD}. \cite{Tingay_2016} and \cite{Tingay_2018} generated spectra towards known exoplanetary systems from the Kepler catalogue \citep{Akeson_2013}. The searches used the same fields and data generated from the Orion region \citep{Tremblay_Orion, Tremblay_RRL} and Galactic plane \citep{Tremblay_2017} surveys to search for molecular and recombination lines with the MWA. However, \cite{Tremblay_2020_SETI} and \cite{Tremblay_2022_SETI} searched the total field of view of 200\,deg$^2$ for single-channel (10\,kHz) radio emission outside of the expected false positive rate across millions of stars and thousands of galaxies \citep{Tremblay_EG}.

Taking a different approach, \cite{Johnson_2023} used the Irish and Swedish LOFAR stations to record observations simultaneously but independently towards the same patch of sky  for a non-interferometric aporach. The authors searched the raw voltage data for coincident radio emission from narrowband, broadband, and pulsating signals. The survey covered a total of 232\,deg$^2$ and more than 1.6\,million stars from the \textit{Gaia} catalog \citep{GAIADR3}. After the data were corrected to the same barycentric reference frame, the multi-site observations provided an effective RFI filtering methodology while the raw voltage data allowed the flexibility to search via multiple methodologies for various transients and technosignatures. 

The RIMS (Radio Interferometric Multiplexed Spectroscopy) approach applies established visibility-domain rephasing techniques in a systematic and multiplexed manner for technosignature and burst searches. This has recently been developed for the LoTSS LOFAR survey \citep{Callingham2025,Tasse2026} and for NenuFAR observations \citep{Zhang2025}. RIMS uses residual visibilities recorded in imaging mode in order to compute dynamic spectra in the four Stokes parameters (I,Q,U,V) at the native time-frequency resolution of the observations toward any point of the field of view simultaneously. It has proved its efficiency by allowing \cite{Callingham2025}, \cite{Tasse2026}, and \cite{Zhang2025} to detect many bursts from stellar systems -- some of which host known exoplanets -- undetected in time-frequency integrated images.

Extending this approach to SKA-Low, enabling comparable analyses will require access to calibrated residual visibilities at native time–frequency resolution, or equivalent multi-phase-centre processing capabilities within standard imaging modes. While the underlying visibility rephasing techniques are well established, their effectiveness for technosignature searches depends on retaining sufficient spectral resolution and temporal fidelity to reconstruct dynamic spectra in all Stokes parameters toward arbitrary sky positions across the primary beam. It will therefore be important that relevant SKA-Low observing modes and data products support such visibility-domain analysis, either through routine archival of residual visibilities or through operational modes that permit flexible beamforming and phase-centre shifting. Ensuring this capability will maximise the scientific return of wide-field technosignature and burst searches with the SKA.

In standard operation, each SKA-Low station beam has a full-width-half maximum between $\sim 9.3^\circ$ at 50\,MHz and $\sim 1.3^\circ$ at 350\,MHz. However, a similar coverage to \cite{Johnson_2023} could be achieved by forming multiple beams, or by using SKA-Low's substation capability. The significant advantage of the SKA-Low for technosignature experiments will come from the large bandwidth across 50--350\,MHz, the point-source resolution, and the sensitivity. For a 1 hour imaging observation toward the Galactic Center using AA*, we can achieve a sensitivity of 22mJy\,beam$^{-1}$ sensitivity at 14\,Hz spectral resolution. Therefore, at a distance of 1\,pc the effective isotropic radiated power (EIRP) limit would be $\sim10^6$\,W, surpassing our most sensitive modern experiments by at least 3 orders of magnitude. More details are provided in Sections 4 and 5.

\subsubsection{Technical Consideration with SKA Low}
As discussed by \cite{Tremblay_PhD}, there are additional technical challenges when performing surveys at low radio frequencies and when using aperture arrays. These challenges need to be considered when using the SKA-Low telescope and include changes in the primary beam due to the large fractional bandwidth, ionospheric effects, and the complexity of the shape of the primary beam and side-lobe emission. 

\subsection{Observation with SKA Mid}
SETI programmes on large dish-based interferometers are already well established at centimeter and millimeter wavelengths, where higher observing frequencies and substantial collecting area provide strong constraints on technosignature radio emission. SKA-Mid will extend these capabilities through its improved sensitivity, wide instantaneous bandwidth, and multiple receiver bands. Although its larger dish size results in a smaller field of view compared to SKA-Low, the higher angular resolution and access to higher frequencies enable more precise localization of candidate signals and tighter limits on equivalent isotropic radiated power (EIRP) from nearby systems.

In practice, technosignature searches at these frequencies typically involve forming coherent beams toward targets within the primary beam of the telescope (see Section 3) and conducting real-time searches for narrowband drifting signals. Existing programmes at facilities such as the Karl G. Jansky Very Large Array \citep{tremblay_cosmic} and MeerKAT \citep{Czech_2021} demonstrate the scientific value of high-frequency, high-resolution interferometric searches. SKA-Mid’s enhanced sensitivity and frequency coverage will allow similar techniques to probe larger volumes of space and place more stringent constraints on potential transmitters.

%SETI programmes on large dish-based interferometers are already well established and take advantage of multicast Ethernet, providing a copy of the antenna data streams. The higher frequencies and larger dish sizes of SKA Mid necessarily imply a smaller field of view than its low frequency counterpart, but offer higher angular resolution than the SKA-low and a wider frequency coverage thanks to multiple receivers. Such systems typically form coherent beams towards targets of interest in the field of view of the primary observing target (See Section 3), and narrow-band drifting signal searches are run in real-time. The primary examples of these systems are Commensal Open-Source Multimode Interferometer Cluster (COSMIC) on the Karl G. Jansky Very Large Array (VLA) \citep{tremblay_cosmic} and the Breakthrough Listen User Supplied equipment (BLUSE) on the MeerKAT telescope (\citealt{Czech_2021}, Czech et al. in prep). 

Given MeerKAT’s role as a precursor instrument forming the core of SKA-Mid, it serves as an ideal template to design future technosignature surveys with the full range. The BLUSE system can form up to 64 coherent beams within MeerKAT’s field of view, selected from a predefined target list, and records data at approximately hertz-level spectral resolution. Although raw data products are not stored due to the prohibitive data rates, a real-time narrowband technosignature search is performed, with “stamp” files—small segments of time–frequency data—saved around promising candidate detections. 

A similar framework could be readily adapted for SKA-Mid, with appropriate consideration to the data ingestion rates. SKA-Mid will provide improved sensitivity, angular resolution, and frequency coverage compared to MeerKAT. Even without commensal data streams, in its standard 1-hour correlator mode, SKA-Mid is expected to achieve a noise level of $\sim$3,mJy,beam$^{-1}$ at 210,Hz spectral resolution in Band~2 under the AA* configuration—approximately three times more sensitive than MeerKAT when extrapolated to a comparable spectral resolution from its 1633,Hz L-band mode. However, due to frequency and time resolution limitations in this mode, a BLUSE-like system will therefore be essential to enable the higher spectral resolutions required for advanced narrow-band technosignature searches. As mentioned above, the storage of candidate data products is already a limitation on commensal techniques and, with the increased data rates expected from SKA-Mid, this may remain a crucial factor in conducting technosignature searches.

A key advantage of interferometric arrays is their ability to distinguish between signals originating in the astronomical far field and those produced by terrestrial or near-field interference. This discrimination—made possible through phase coherence across widely separated antennas—is essential for confirming whether a candidate signal is truly astrophysical. In addition to traditional beamformed searches, interferometric imaging provides a powerful complementary tool for this purpose. By forming images of candidate detections (``postage stamps”), it is possible to localize emission on the sky and examine its spatial structure, helping to determine whether it is consistent with a distant celestial source or with local interference. The increased sensitivity and angular resolution of SKA-Mid will significantly enhance this capability, strengthening its role in the confirmation and characterization of potential technosignature candidates.

%In addition to beamformed observations, where technosignature searches are typically conducted, the image plane is also a powerful space for identifying if a candidate SETI signal is sky-bound, and imaging of stamp files to constrain signal morphology is becoming more common. The heightened sensitivity and resolution of SKA-Mid significantly increase the potential of this technique, which will form an important aspect of any confirmation process for a candidate technosignature.

Beyond candidate follow-up, the image plane is also being expanded as a discovery space in its own right. The discovery of long period radio transients \citep[e.g.][]{2022Natur.601..526H}, pulsators with periods of tens to thousands of seconds, has ignited an interest in fast (correlator integration time) imaging across the considerable field of view of arrays such as MeerKAT \citep[e.g., TRON;][]{Smirnov_2024,Smirnov_2025,Smirnov_2025_2}, the Australian SKA Pathfinder, and LOFAR. These searches attempt to identify sources that have a varying flux density within an observation or appear in only a subset of images. As these pipelines mature, the high spectral resolution and polarisation axes are also being explored to find sources with unusually high polarisation fractions or unusual spectral shapes \citep{2023MNRAS.525L..76H}. When applied to technosignature candidate identification, these techniques move beyond the standard narrow-band drifting signal searches and are somewhat agnostic to source temporal and spectral properties (albeit not probing Hz level spectral resolution). This technique is particularly exciting given that fast imaging is one of the current observatory-delivered data products, and so routine SETI surveys could be performed with this mode. Fast imaging can also provide important ancillary information for more traditional technosignature searches, for example, identifying periods where satellite(s) are passing through the primary beam. However, the advent of satellite megaconstellations may present an accelerating source of false positives to technosignature searches \citep[e.g.][]{Grigg2025}. Nevertheless, advances in the efficiency of these fast-imaging pipelines open up the possibility of forming coarse (a few seconds, tens of kHz) dynamic spectra over the entire field-of-view, opening up new avenues for machine learning driven anomaly detection. 

As discussed in the SKA-Low section (Section 2.1), the RIMS technique can be applied to SKA Mid data as well.

\section{Target Selection}
\label{ssec:target_selection}

As we do not know in advance which direction on the sky might host a technosignature, the ideal experiment would continuously monitor the entire sky over a wide frequency range. In practice, however, even with modern computing advances, true all-sky, all-time searches remain limited by data rates, storage, and processing capacity. For facilities such as the SKA, a more tractable strategy is therefore to observe large areas of sky while prioritizing predefined source catalogues and processing only the relevant phase centres in detail.

A practical approach for the SKA would begin with a master input catalogue—such as the 30 million stars identified in the \textit{Gaia} survey \citep{Czech_2021}—each with known sky coordinates and distances. During standard survey or targeted observations, the telescope field of view would typically contain hundreds to thousands of these catalogue entries at any given time. Rather than slewing to each of 30 million targets individually, the SKA would observe wide fields and digitally form coherent beams toward all catalogue sources within the primary beam. In this way, millions of targets can be searched opportunistically during routine observations, dramatically increasing survey efficiency.

Target preselection also serves practical purposes beyond reducing the search space. Known exoplanet hosts, nearby stars, ultracool dwarfs, and other well-characterized systems provide physically motivated targets and enable more informed signal filtering. For example, when orbital parameters are known, constraints can be placed on expected Doppler drift rates, transit times, or occultation geometries, which can help distinguish potential astrophysical signals from terrestrial interference \citep{Sheikh_2019,Li_2022,Li_2023} While nearby stars remain attractive targets because sensitivity limits translate directly into stronger constraints on transmitter power, wide-field interferometers like the SKA are not restricted to single-object observations and can simultaneously survey large stellar populations.

In addition to star catalogues, expanded target lists now include exotic or non-traditional systems such as galaxies, H II regions, and other astrophysical environments \citep{Lacki_2021}. Modern techniques can synthesize dynamic spectra toward every object in an input catalogue within the field of view, allowing for scalable searches limited primarily by compute resources rather than telescope pointing time. This approach has already been demonstrated on current commensal backends \citep{tremblay_cosmic,Czech_2021} would scale naturally to the SKA, whose collecting area and digital infrastructure are designed for high-throughput survey science.

Geographic location must also be considered. Both SKA-Mid and SKA-Low are located in the Southern Hemisphere, which provides excellent access to southern sky targets but introduces a declination bias relative to northern facilities. However, because the southern sky contains a large fraction of the Milky Way plane and nearby stellar populations, this bias does not significantly limit large-catalogue searches. For example, out of the 33\,million stars in \cite{Czech_2021}, 26\,million are observable with MeerKAT.

Finally, while most technosignature efforts have historically focused on targeted stars, wide-field interferometers also enable background searches across entire fields. In practice, this means that every observation can simultaneously constrain emission from any object within the field of view, not only the catalogued targets. For the SKA, the combination of large field of view, digital beamforming, and high-performance computing would allow the systematic processing of tens of millions of predefined targets over time—without requiring tens of millions of individual pointings. This could open opportunities for observing additional targets like ``Schelling points'': locations that may be mutually-derivable to the transmitter and the observer. (\cite{Wright_2017}).

\section{Observation Strategy \& Signal Processing}
Interferometry has led to a revolution in SETI experiments, allowing for increases in sensitivity, localisation, and potential false positive discrimination. Modern interferometric SETI is performed through both dedicated observing time and commensal campaigns on arrays. These interferometric SETI surveys currently rely primarily on incoherent or coherent beamforming modes, particularly modes with a high spectral resolution data product (with typical channel widths of the order $1\,\rm{Hz}$). Coherent beamforming, in particular, has the advantage of providing localisation information for any candidate signal, but with the disadvantage of needing to form many coherent beams to tile the entire field of view (as would be optimal for searching for technosignature candidates). Coherent beamforming is also a computationally expensive operation, and it is usually not feasible to save the beamformed data for every source within the field. Usually, the beamforming approach used in SETI strategies involves prioritising nearby stars or known exoplanet systems (as discussed in Section~\ref{ssec:target_selection}) based on optical catalogues such as those from \textit{Gaia}. The number of viable targets depends strongly on the observing frequency (with a reduced field of view at higher frequencies), but less so on the pointing direction, as nearby stars are mostly isotropically distributed across the sky. 

Alternately, some new surveys are exploring the use of correlator backends for technosignature surveys, or ``image-plane SETI'', in which images are made from corrected visibility products are formed between pairs of antennas (e.g. \citealt{Tingay_2016,Tingay_2018}). %These visibilities can then be related to the sky brightness distribution observed through the van Cittert–Zernike theorem. T
This allows the experiment to take full advantage of the interferometer's ability to have point source location information and sensitivity across the full field without having to preselect targets of interest. This style of observation would closely mimic image-based pulsar searches and spectral line experiments (e.g., \citealt{Cotton_2016,Smirnov_2025}). However, image-plane SETI requires advanced and subtle processing techniques: for example, it is necessary to perform reference frame correction and modelling of the Doppler signal drift, as the signal could span multiple frequency channels (when channels are less than 1\,kHz, \citealt{Mason_2025}). Additionally, creating image products at high temporal and/or spectral resolution for small dish long baseline arrays across the entire observing band is computationally challenging. ``Zoom modes'' or technosignature searches at lower spectral resolution when compared to beamformed outputs are required to take advantage of the full field of view in this way.   

The imaging zoom mode offered in AA* of the SKA-Low offers up to a 14\,Hz resolution with a corresponding 24\,kHz bandwidth per zoom window, as the maximum instantaneous bandwidth scales with the channel resolution. For SKA-Mid the highest resolution zoom mode has 210\,Hz resolution over 3.125\,MHz of bandwidth. Lower resolution zoom modes that cover more of the accessible bandwidth are available for both instruments. These SKA modes would allow for narrowband or spectral line width searches at various spectral resolutions, periodicity or second-scale transient searches, or machine learning approaches on dynamic spectra or saved image data. SKA correlator data could also be used to follow-up or localise signals detected from other instruments, adding to the suite of instruments used in the search for life beyond Earth.

Technosignature searches in the image plane, or using correlation data products, are currently relatively under-utilised \citep[although see e.g.][]{Tremblay_2020_SETI,Wandia_2023,Mason_2025}. An important component of commensal SETI experiments (such as COSMIC and BLUSE) is the ability to store raw antenna voltages when a signal of interest is detected in a beamformed data product. These so-called stamp files can then be correlated and imaged post-observation, and a genuine technosignature candidate should have the properties of a point source (see also Section 7) as opposed to the contaminant RFI that will typically appear in the near field \citep{prabu_2023}. While beamformed and single-dish technosignature searches typically focus on narrow-band (Hz level) drifting signals, general spectro-temporal searches across the entire field of view can be implemented in the image plane in order to identify (and localise) unusual astrophysical sources (e.g., \citealt{Tremblay_2020_SETI}). 
%An intermediate step to a direct image-plane search is to search the correlator properties directly by forming dynamic spectra from clean residuals across the field of view. Once the static sky has been subtracted, the visibilities can be re-phased to arbitrary positions on the sky (limited by the field of view of the instrument) and spectro-temporal information can be measured at the native correlator resolution. While this is computationally expensive, recent software developments have made it feasible to create dynamic spectra over large fractions of the field of view (see e.g. \url{https://github.com/saopicc/RIMS}). Such a data product lends itself well to anomaly detection techniques.

An efficient intermediate step to a direct image-plane search or coherent beamforming is to form dynamic spectra from clean residual visibilities across the field of view. Once the static sky has been subtracted, the visibilities can be re-phased to arbitrary positions on the sky (limited by the field of view of the instrument) and spectro-temporal information can be measured in four Stokes at the native correlator resolution. Recent software developments have made it feasible to create hundreds of thousands of dynamic spectra over large fractions of the field of view at a modest computational cost of a few percent of the imaging process (see and \url{https://github.com/saopicc/RIMS}). This data product lends itself well to anomaly detection techniques \citep{Tasse2026,Callingham2025,Zhang2025}. As the synthesis of dynamic spectra uses the residual visibilities from the imaging process, it must be performed at the SDP level where these products are still available.

The pulsar backend on SKA would allow for a different and complementary set of SETI investigations. The coherent beamformed mode has coarser frequency resolution and lower sensitivity (due to the reduction of the number of antennas allowed while using this mode), but would allow us to search for periodic spectral emission at kHz or greater scales (e.g.,\citealt{Suresh_2023}). This kind of signal morphology may even be more efficient than narrowband signals for transmitters that broadcast for long periods of time \citep{Gajjar_2022} that we are statistically most likely to detect. This is discussed in more detail in Section 4.1.

\subsection{AA* Output Level Data Products}
\label{AA*}
The Science Data Processor (SDP) systems for SKA-Low and SKA-Mid will produce a number of calibrated ``output level data products'' \citep[OLDPs, see][]{skao_data_products_2025}, upon which we may conduct technosignature searches. The OLDPs are standardised across SKA-Low and SKA-Mid, and form the basis of what searches are possible with AA*.

\subsubsection{Continuum mode}
Continuum observing mode will enable wide-bandwidth technosignature searches over large sky volumes, but with lower sensitivity than that achievable with higher resolution data products. When observing in continuum mode, the SKA-Low and Mid telescopes have frequency resolutions of 5.4\,kHz and 13.4\,kHz, respectively. Although this resolution is not high enough to perform Doppler drift searches, narrowband technosignatures and anomalous spectral features may still be detected following a strategy similar to \cite{Tremblay_2020_SETI}. Searches could be conducted on either the spectral-line image cube OLDP or in cubes/dynamic spectra generated in post-processing from the calibrated visibilities OLDP. 

Technosignature searches in continuum data products will have significantly lower sensitivity than searches in higher resolution data products, if we assume a Voyager-like transmitter signal. The radiometer noise within each frequency channel increases as $\sqrt{\Delta\nu}$ with increasing channel bandwidth; as such, narrowband technosignature searches at 5.4\,kHz resolution will have 73$\times$ less sensitivity than a search conducted on 1\,Hz resolution data, and searches on  13.4\,kHz product would have  116$\times$ lower sensitivity for a putative narrowband technosignature with $\leq 1$\,Hz bandwidth. However, if a signal is naturally a broad-band emitter we would expect similar sensitivity to a spectral line experiment.

\subsubsection{Zoom mode}
Order-of-magnitude more sensitive narrowband searches are possible by leveraging the higher frequency resolution offered by `zoom' observing modes; however, hardware constraints and data volumes limit the maximum number of frequency channels that can be processed at once. The SKA-Low and SKA-Mid zoom modes offer up to 14\,Hz and 210\,Hz resolution, respectively. For AA*, it is anticipated that SDP can deliver up to 64,000 zoom channels within an image cube, which would place the maximum processed bandwidth for SKA-Low for 14\,Hz channels at  0.9\,MHz, and for SKA-Mid, a 13.4\,MHz bandwidth for 210\,Hz zoom channels.

Narrowband technosignature searches on zoom mode data would need to employ de-Doppler algorithms to recover drifting signals. As time averaging will be required to keep data volumes tractable, a loss in sensitivity due to time smearing is expected regardless of OLDP. Following \cite{Sheikh_2019} a fractional Doppler drift rate of 2 nHz is sufficient for terrestrial planets; this corresponds to 0.1--0.7\,Hz/s across the 50--350 MHz SKA-Low band, and 0.7--3.08 Hz/s across the 0.35--15.4\,GHz SKA-Mid band. To search a wider class of exoplanetary hosts, \cite{Sheikh_2019} recommend that Doppler drift rates of up to 200 nHz should be searched (i.e. 100$\times$ larger). 

\subsubsection{Pulsar timing: filterbank and flowthrough modes}

For targeted technosignature searches of nearby stars and exoplanets, coherent `tied-array' beams can be formed using the the pulsar timing subsystem (PST); in AA*, SKA-Low will form up to 8 tied-array beams (300 MHz bandwidth / beam), and SKA-Mid will form up to 16 tied-array beams (2.5 GHz bandwidth / beam). However, the tied-array beams only include stations within a 10\,km core radius, so do not reach the full array sensitivity. Although the pulsar timing solution output of PST is of little technosignature interest, PST will also provide three OLDPs that are highly relevant to technosignature searches: 

\paragraph{Detected filterbank archives} 

The detected filterbank processing mode generates dynamic power spectra with configurable time and frequency resolution (with maximum resolution of 100 ns and 2 Hz, respectively), for all four pseudo-Stokes products (I,Q,U,V). This data product can support a wide variety of technosignature search strategies. 

For narrowband technosignature searches, the 2\,Hz resolution will result in improved sensitivity than the 14\,Hz (SKA-Low) and 210\,Hz (SKA-Mid) zoom modes due to less radiometer noise within each channel---despite the tied-array beam only including stations within a 10\,km radius. Sensitivity loss due to time smearing can also be minimized by using higher time resolution than that possible for an image cube. Alternatively, by forming dynamic spectra with high time resolution, searches for pulsed or periodic spectral emission at kHz or greater scales may be conducted.

\paragraph{Flow-through and VLBI modes}

The PST mode can also generate a `flowthrough' OLDP that contains a decimated version of the tied-array beam voltage data, stored with a bitdepth of 1,2,4 or 8-bits. This voltage-level OLDP affords the most flexibility for technosignature searches and could be post-processed to run searches for multiple different classes of technosignatures. 

The beam voltage data can also be output as VLBI data products in VDIF format ready for use in VLBI observations; see Section \,\ref{sec:vlbi} for further details on VLBI technosignature searches.

\subsubsection{Pulsar search mode}

Unlike PST, the pulsar search subsystem (PSS) does not provide direct access to dynamic spectra. Instead, the primary OLDP is a list of sieved pulsar and transient candidates formed from a pulsar acceleration search pipeline and a single-pulse search. While these OLDPs are tuned for pulsar and fast radio burst science, the single-pulse search pipeline does deliver dynamic spectra around candidate de-dispersed signals, upon which a search for anomalous signals could be conducted. In general, we do not expect the PSS data products available in AA* to have broad application for technosignature searches.

\section{Sensitivity}
\begin{figure}
    \centering
    \includegraphics[width=\linewidth]{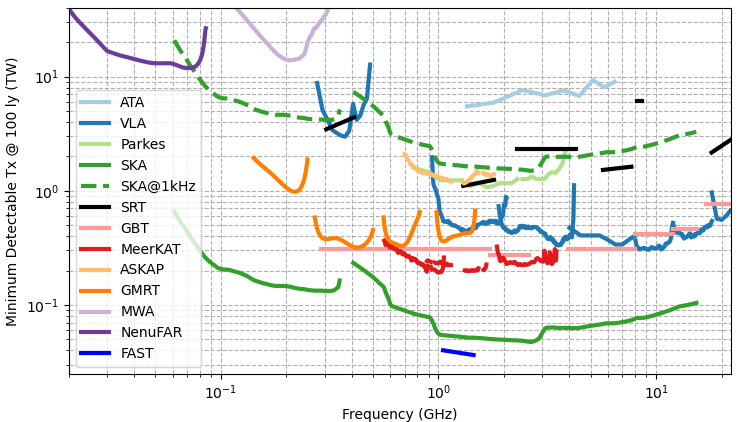}
    \caption{Detectable narrowband transmitter at 100 ly in terawatt (TW)} assuming a signal-to-noise ratio of 5 and 5 minutes integration.  The bandwidth is 1 Hz for all, but to illustrate the need for narrowband processing the SKA at 1 kHz has been included in the dashed green line.\label{fig:sensitivity}
\end{figure}

The SKA will provide a great improvement in raw sensitivity across the band.  However, in order to realise the full sensitivity appropriate for technosignature searches, the system must be able to flexibly process appropriate bandwidths and timespans. The narrowband detection performance of various radio telescopes is shown in Figure \ref{fig:sensitivity}, most of which have active technosignature search programmes.  The figure scales for the effective isotropic radiated power (EIRP) detectable at 100 ly with a signal-to-noise ratio of 5 with a 5 minute de-drifted integration time. The bandwidth is 1 Hz, except for the curve labelled ``SKA@1kHz'', which demonstrates the impact of not being able to appropriately process narrow enough bands.

\section{Search Algorithms \& Strategies}
\citet{Wright_2022} note that there is no fundamental reason why technosignatures could not be more abundant, longer-lived, or more detectable than biosignatures. To date, most technosignature searches have relied on matched-filter approaches, in which algorithms are designed to detect specific classes of anticipated signals (e.g., narrowband drifting tones). While this strategy remains central to current efforts, advances in digital signal processing and high-performance computing—driven by sustained improvements in hardware capability and software efficiency—have expanded the range of analyses that are computationally feasible. In particular, GPU-accelerated processing, scalable data pipelines, and machine learning–based classification tools now allow for broader, less assumption-dependent searches within defined observational constraints. These developments do not replace targeted search strategies, but they enable complementary approaches that reduce reliance on strictly predefined signal models.

%\cite{Wright_2022} argue that ``there is no incontrovertible reason why technology could not be more abundant, longer-lived, more detectable, and less ambiguous than biosignatures.'' 
%Although most searches for technosignatures consist of matched filters -- designing an algorithm to search for what we expect to find -- we have reached an era where we can implement more agnostic search methodologies as well; the rise in agnostic search methodologies can be attributed to significant improvements in GPU-accelerated processing, efficient coding practices, and the use of artificial intelligence algorithms instead of manual data perusal. 

\cite{Sheikh_2019} discuss a number of search algorithms for Doppler drifted narrowband signals and compare them with respect to efficiency and results. The authors argue that the Taylor-Tree de-Doppler algorithm is the most well suited for modern searches. This algorithm is implemented in \textsc{TurboSETI} \citep{Enriquez_2019TT}\footnote{\url{https://github.com/UCBerkeleySETI/turbo_seti}} and \textsc{seticore}\footnote{\url{https://github.com/lacker/seticore}} (a GPU accelerated version of the \textsc{TurboSETI} software). Another approach uses a kurtosis-based statistical algorithm. This was implemented in a software package called \textsc{hyperseti}\footnote{\url{https://hyperseti.readthedocs.io/en/latest/}} for fast transient and technosignature searches and in a pipeline called \textsc{pickles}\footnote{\url{https://github.com/UCBerkeleySETI/pickles}}.  \cite{Painter_2024} used this method to search through observations of 2841 stars across four frequency bands in a highly efficient way. 

As more SETI experiments are conducted using interferometric arrays, including the SKA, new techniques are being developed to improve the separation of radio frequency interference (RFI) from genuine astronomical signals\footnote{We note here that this is particularly important in SETI, as discussion in Section 1, the signal we are looking resembles closely to RFI.}. For example, \citet{Tusay2024} used the Allen Telescope Array to form two coherent beams simultaneously--one centred on the target source and another placed at the half-power point of the primary beam--and performed a cross-correlation analysis to reject signals present in both beams, which are likely RFI. Similarly, \citet{Jacobson-Bell_2025} applied an AI-based clustering algorithm to single-dish data, using spatial filtering through differential pointing. In their method, the algorithm was trained to identify signals appearing only when the telescope was pointed at the target and absent when pointed at a nearby ``control'' sky position.

Self-supervised and semi-unsupervised deep learning approaches have also shown promise for RFI rejection and signal-of-interest detection \citep{self-supervised-seti, ma_ts_deeplearning}, which could be of interest to other science fields using the SKA (e.g. \citealt{Yang_2025}). These methods employ autoencoders to generate latent-space representations of spectrogram data, which can be used to reconstruct or predict waterfall plots. This enables the models to learn RFI morphologies and distinguish them from narrowband candidate signals. \citet{self-supervised-seti} utilised a stacked convolutional long short-term memory (ConvLSTM) network for temporal prediction and spatial filtering, comparing reconstructed on-target waterfall plots with corresponding control observations. In contrast, \citet{ma_ts_deeplearning} trained a $\beta$-Variational Autoencoder ($\beta$-VAE) on Green Bank Telescope data with synthetic signal injections produced using the software \textsc{setigen} \citep{setigen}. They performed spatial filtering by measuring the distance between encoded feature vectors from on-target and control observations, under the assumption that genuine technosignatures should yield highly dissimilar encodings. Notably, \citet{ma_ts_deeplearning} identified eight previously undetected signals of interest and reported a lower false-positive rate compared to the Taylor-tree method implemented in TurboSETI on the same dataset.

As discussed in Section 1, not all potentially detectable emission from technology follows the spatial-filtered, narrowband structure. \cite{Gajjar_2022} created a GPU-accelerated search pipeline called \textsc{SPANDAK} to find both astrophysical broadband signals with a characteristic dispersion sweep, but also signals with time- or frequency-inverted sweeps that would indicate an artificial origin. Similarly, \cite{Suresh_2023} designed a fast folding algorithm to look for pulsating signals limited by frequency in radio astronomy datasets. Both of these have been tested on data from the Green Bank telescope, but no anomalous signals were detected. 

To date, most broadband searches have been performed in the image plane (e.g., \citealt{Tingay_2016}), which offers a lot of flexibility over the traditional single-pixel ``waterfall plot'' (dynamic spectra) searches that dominate the field. In particular, leveraging the entire field of view using interferometric data provides high sensitivity, arcsecond point source resolution, and location-agnostic search design. 
As an alternative, the RIMS technique will make these much more efficient, multiplexing the time-frequency searches simultaneously over many pixels across the field of view.

%\begin{table}
%\centering
%\begin{tabular}{|l|c|c|c|r|}
%\hline
%\textbf{Experiment} & \textbf{Frequency} & \textbf{Program} &\textbf{EIRP$_{min}$} \\
% & \textbf{GHz} &  & \textbf{W} \\
%\hline
%COSMIC-VLA & 2--50 & Commensal & $2\times10^{12}$ \\
%\hline
%BLUSE- MeerKAT & 0.58 --4 & Commensal & $1\times10^{12}$ \\
%\hline
%%ASKAP & 0.74--5 & Guest Science&  Data 8 \\
%%\hline
%MWA & 0.08--0.3 & Allocated Time& $3.2\times10^{13}$ \\
%\hline
%Allen Telescope Array & 1--12 & Dedicated Time &  $4\times10^{14}$ \\
%\hline
%%GMRT & 0.15--1.5 & Dedicated Time &Data 8 \\
%%\hline
%%%\hline
%Parkes 64m & 0.7--4 & Dedicated Time & $1\times10^{13}$ \\
%\hline
%GBT & 1--8 & Dedicated Time & $2\times10^{12}$ \\
%\hline
%SRT & 4--12 & Dedicated Time & $8\times10^{17}$ \\
%\hline
%\end{tabular}
%\caption{Table of current radio SETI projects on precursor and pathfinder telescopes. The %EIRP$_{min}$ is the realized or projected transmitter data rate.}
%\label{tab:example}
%\end{table}

\section{Commensality}

\begin{figure}
\includegraphics[width=0.85\textwidth]{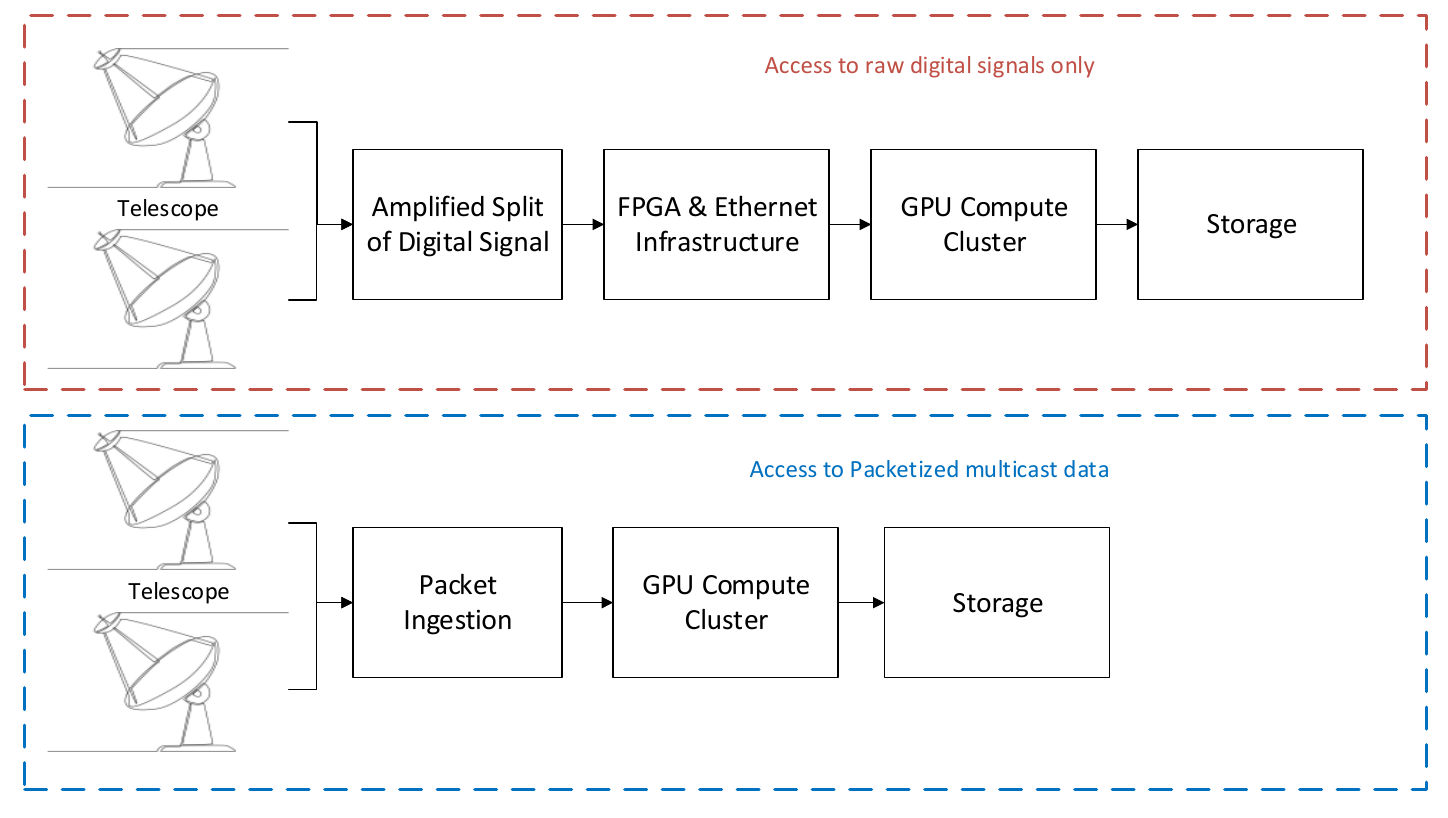}
\caption{Two potential digital architectures for commensal systems. Top: a system which provides a copy of the digital signals in a raw form where the commensal system would include a signal processing chain. Bottom: the architecture if multi-cast data is provided in the form of calibrated coarse channels. Both are viable options but it is more cost-effective if the observatory provides a multi-cast setup.  \label{fig:Commensal}}
\end{figure}

At radio wavelengths, it is relatively straight-forward (and scientifically viable because of the many photons per state of the system) to make multiple copies of a received bandwidth and process them independently for different science cases, as demonstrated in Figure \ref{fig:Commensal}. This technical flexibility makes simultaneous observing modes both practical and efficient, particularly for high-risk, exploratory programmes such as technosignature searches. Since we do not know a priori what form a successful detection might take, or even what constitutes an unambiguous signal of extraterrestrial technology, the ability to run parallel experiments on the same telescope provides a significant scientific advantage without requiring additional dedicated observing time.

Searching the Galaxy for evidence of technologically active species remains an enormous undertaking. Although radio technosignature searches are beginning to place quantitative constraints on portions of parameter space through null results, much of this progress has occurred only in recent years \citep{Wright_2018,Tremblay_2025}. Fully characterizing the distribution of technologically capable life would require substantially more observing time than can realistically be allocated under a traditional single–primary-user model. Commensal observing—attaching a secondary backend to an existing telescope system—provides a practical solution by increasing observational coverage while minimizing additional operational cost.

Historically referred to as “parasitic SETI” \citep{Bowyer_1983}, the field now adopts the term “commensal SETI,” borrowing the biological concept in which one organism benefits without harming the host. In astronomy, commensal operation refers to accessing raw or duplicated data streams from a telescope without interfering with the primary science programme. While archival data analyses can address some technosignature questions (e.g., \citealt{Cohen_1980,Mason_2025}), dedicated commensal systems offer an important advantage: they can generate optimized data products tailored to technosignature searches, including high time- and frequency-resolution streams not typically preserved in standard survey archives.

%Although Jill Tarter may not have used the specific term commensal in her published literature, the principles underlying commensal SETI were embodied in her pioneering work on projects such as SERENDIP—where SETI data were collected alongside primary astronomical observations—and in her advocacy for shared, real-time access to telescope data

Although Jill Tarter may not have used the specific term commensal in her published literature, the principles underlying commensal SETI were embodied in her pioneering work. The importance of simultaneous observations was also emphasized in early strategic studies such as Project Cyclops and the SETI 2020 Science Working Group \citep{SETI2020}, which identified continued growth in computational capability as a key enabler for increasingly sensitive and specialized searches. Early implementations of this approach include the SERENDIP programme \citep{Bowyer_1983,Werthimer_2001}, which conducted commensal experiments at Hat Creek, Arecibo, and the Green Bank Telescope across multiple frequency bands. SERENDIP reported hundreds of anomalous signals \citep{donnelly_serendip_1998}, illustrating the scale and discovery potential of continuous, wide-coverage observing. Additional distributed and commensal efforts at Arecibo included SETI@home and AstroPulse \citep{2013ApJ...767...40V,2025AJ....170..112K,2025AJ....170..111A}. These efforts reflect the underlying philosophy of maximizing observational efficiency and enabling multiple science cases from the same data stream.

Modern radio facilities have further expanded these capabilities. Commensal technosignature systems now operate at the Karl G. Jansky Very Large Array \citep{tremblay_cosmic}, MeerKAT (Czech et al. in prep; \citealt{Czech_2021}), and the Murchison Widefield Array \citep{Morrison_2023}. MeerKAT, in particular, was among the first large facilities to adopt a multicast Ethernet architecture \citep{SGP_2022}, enabling real-time duplication and distribution of data products to independent processing nodes. This design allows multiple scientific programmes to operate concurrently on the same incoming data stream. The broader scientific value of such architectures is demonstrated by commensal discoveries beyond technosignatures, including fast radio bursts and other transient phenomena \citep{Caleb_2020,DLN_2020}.

%Together, these developments illustrate how commensal observing leverages advances in digital signal transport and high-performance computing to expand sky coverage efficiently. For large facilities such as the SKA and its precursors, this model provides a scalable path toward sustained, wide-parameter-space technosignature searches without competing directly with primary observing programmes.

Modern commensal systems rely on well-defined digital architectures that allow duplication and redistribution of telescope data streams with minimal impact on primary operations (Figure~\ref{fig:Commensal}). Current implementations generally follow one of two models:

\begin{itemize}
\item The observatory provides raw digitized voltage streams, and the commensal system performs all downstream transformations, including F-engine processing and higher-level computation;
\item The observatory provides coarse-channelized and calibrated data products via an Ethernet multicast architecture following the F-engines, enabling downstream processing without duplicating early-stage signal conditioning.
\end{itemize}

Experience from existing systems—such as COSMIC at the VLA and BLUSE at MeerKAT—demonstrates that both approaches are technically viable. However, the second model offers greater scalability, as multicast architectures allow multiple independent commensal projects to subscribe to shared data products with minimal additional overhead. This framework enables substantial increases in scientific output without requiring significant modifications to the primary observing infrastructure.

In addition, techniques such as RIMS \citep{Tasse2026} extend the concept of commensality further by operating directly on residual interferometric visibilities. RIMS builds on standard imaging pipelines, it can be parallelized efficiently and increases computational costs by only 1–2\%, demonstrating that commensal technosignature searches can be integrated into standard interferometric workflows with negligible additional burden.

\section{VLBI Considerations}

\label{sec:vlbi}
\begin{figure}
\includegraphics[width=1.0\textwidth]{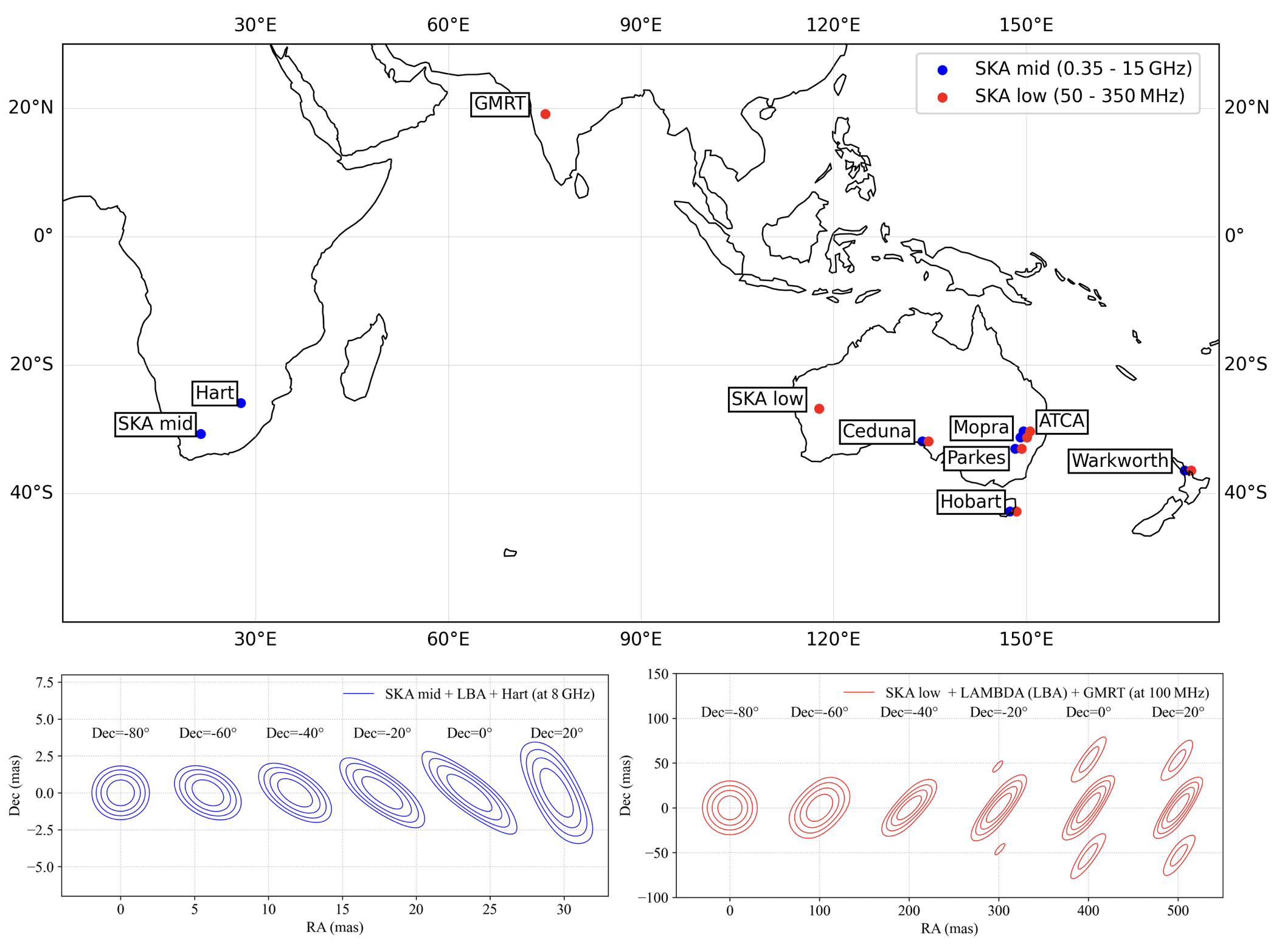}
\caption{Top: deployment of potential VLBI stations for SKA-Mid and SKA-Low. The locations of existing LBA stations are adopted as candidate sites for potential low-frequency VLBI facilities (e.g., LAMBDA) associated with SKA-Low. Bottom: Synthesized beams for VLBI observations with SKA-Mid and SKA-Low at different source declinations.}
\label{fig:stations}
\end{figure}

The SKA is expected to support Very Long Baseline Interferometry (VLBI) capability for both SKA-Mid and SKA-Low. This will be an important asset for technosignature studies, especially for the follow-up of interesting candidate signals. Figure \ref{fig:stations} shows the deployment of potential VLBI sites in the Southern Hemisphere for SKA-Mid and SKA-Low, along with the expected synthesised beam for a range of target source declinations. Across the frequency range of 350\,MHz to 15\,GHz, VLBI observations with the SKA-Mid will offer angular resolutions ranging from sub-milliarcsecond to several tens of milliarcseconds. This will enable highly precise localisation of potential SETI signals: for example, at a distance of $\sim 100$~pc, 1~AU subtends an angle of $\sim$10~milliarcseconds. The positions of nearby stars measured by Gaia have sub-milliarcsecond accuracies \citep{2021ARA&A..59...59B}. If a technosignature originates from an exoplanet, VLBI observations can be used to measure the angular offset between the host star and the exoplanet. In addition, multi-epoch VLBI observations separated by months to years would enable the measurement of the proper motion of the technosignature, which is expected to trace the orbital motion of the exoplanet. This opens the possibility of associating a detected technosignature directly with a planetary system, a spacecraft, or other astrophysical structure \citep{Garrett_2018}.

The first VLBI experiment explicitly designed for SETI was conducted by \cite{Rampadarath_2012} using the long baseline array (LBA). They noted that one of the main advantages of VLBI for SETI is the suppression of terrestrial radio frequency interference (RFI). On long baselines, strong RFI typically decorrelates because of the large residual fringe rate for signals located outside of the small correlated field of view. This property makes VLBI a uniquely powerful tool for discriminating between anthropogenic RFI and genuine celestial signals. Some of the ways in which a VLBI array can be used to discriminate technosignature candidates from RFI are,

\begin{itemize}
    \item \textbf{Cross-correlation products:} individual antennas in a VLBI baseline pair are very far apart and see very different RFI. As a result, their cross-correlation products (i.e., visibilities) effectively suppress localised RFI while preserving the common astronomical signal, such as a technosignature. The decoherence factor characterises the degree of correlation of the detected signals between different VLBI stations. Localised RFI sources or signals from moving satellites on the sky exhibit decreasing coherence with increasing integration time, leading to a reduced signal-to-noise ratio in the visibility data.

    \item \textbf{Tracking phase changes:} Tracking changes in visibility phases with time and frequency can also help separate technosignature signals from RFI \citep{Rampadarath_2012}. The technique essentially uses the fact that most terrestrial RFI originates near the horizon (i.e., at very large zenith angles) and therefore exhibits phase (or delay) properties as a function of time and frequency that are markedly different from those of astronomical signals, which typically occur at smaller zenith angles. In addition, RFI generated by satellites outside the Earth’s atmosphere also exhibits distinct characteristics in the fringe rate, as such sources move across the sky much faster than astronomical signals.

    \item \textbf{Near-field techniques:} Although the cross-correlation process suppresses much of the coherence from nearby RFI signals, very strong RFI can still remain. In such cases, the distance to the interfering source can be estimated using near-field imaging techniques. In principle, radiation wavefronts from near-field objects are not parallel, leading to different arrival times at different VLBI stations. Near-field imaging accounts for this effect by applying varying focal distances, thereby enabling distance measurements of radio sources such as orbiting satellites \citep{prabu_2023}.
    
   % the distance upto which the near-field of an interferometer extended scales with the baseline length, which for an VLBI like array encompasses most RFI sources (terrestrial and orbiting) within the array's near-field. Hence, near-field techniques can be used to identify and `peel' RFI signal contributions from the visibility matrix 

    \item \textbf{Range-Doppler analysis:} metrics such as radar ambiguity functions can be used to differentiate satellite based RFI from technosignature signals, as they are both expected to show very different Doppler signatures. 
\end{itemize}

In addition to its RFI rejection capabilities, VLBI offers several other advantages for technosignature searches:

\begin{itemize}
    \item \textbf{Verification through redundancy:} Independent detections of a candidate signal on multiple baselines add robustness that is simply not available to single-dish or beam-formed systems. This is especially important for faint or intermittent signals, where confidence in the astrophysical origin must be maximised \citep{Wandia_2023}.
    
    \item \textbf{Imaging capability:} Interferometer data can be transformed into the image domain across the full primary beam, enabling wide-field searches. During a typical few-hour VLBI observing session, a genuine technosignature should maintain a fixed two-dimensional position on the sky, even if other characteristics such as frequency, temporal modulation, or dispersion vary. This positional invariance serves as a powerful discriminator against false positives.
    
    \item \textbf{Localization through VLBI astrometry:} By cross-matching VLBI-localized signals with high-precision astrometric catalogues (e.g., \textit{Gaia}), it is possible to test their association with nearby stellar systems. Linking a candidate signal to a specific star or exoplanetary system would mark a major advance in the field.
At typical SKA–VLBI resolutions, sub-AU orbital scales around nearby stars can be resolved. If a signal originates from a planet orbiting such a star, its proper motion can be measured through multi-epoch VLBI observations, offering strong evidence for an artificial origin.
\end{itemize}

The ability to combine the SKA with existing global VLBI networks will further enhance these capabilities. For example, SKA-Mid tied into the European VLBI Network (EVN), the LBA, or an African VLBI Network (AVN) would deliver sub-milliarcsecond angular resolution with excellent sensitivity. SKA-Low in VLBI mode, though more technically challenging due to ionospheric effects, would probe complementary frequency ranges, enabling multi-band confirmation of candidate events. In particular, the substantial improvement in baseline sensitivity provided by the SKA is expected not only to broaden the scope of detectable extraterrestrial signals, but also to enable more sophisticated analyses, including the detection of wide-band signals in addition to conventional narrow-band emissions, as well as detailed investigations of their modulation characteristics and signal structures.

Compared to standard interferometric observations, VLBI offers significantly higher angular resolution, making targeted observations of specific sources more effective than wide-area blind surveys. Recent advances in exoplanet research have led to the identification of numerous systems with high habitability potential. Targeted VLBI SETI observations toward these carefully selected systems therefore represent a highly efficient strategy that can greatly enhance the probability of detecting extraterrestrial signals. A further promising application of VLBI is in post-detection verification. In the event of a technosignature candidate identified by SKA in stand-alone mode, rapid-response VLBI follow-up would provide an independent test of its authenticity. Confirming a signal on baselines spanning thousands of kilometres would virtually eliminate the possibility of local RFI contamination. This rapid verification capability is likely to form a cornerstone of any future SETI detection protocol.

Finally, VLBI can also be exploited in a commensal fashion. Modern VLBI networks often observe with wide instantaneous bandwidths, generating large visibility data sets. Searching these data for anomalous broadband signals (especially in fields containing nearby stars) is an efficient way to leverage existing infrastructure for technosignature science, much as \cite{Mason_2025} demonstrated with ALMA in a non-VLBI context. Artificial broadband signals should exhibit high brightness temperatures, making them distinguishable from natural astrophysical phenomena. However, VLBI resolution would be required to ensure that the emission is not associated with a stellar host, thereby eliminating the possibility of misinterpreting the signal as natural in origin. In this sense, VLBI with the SKA is not only a follow-up tool but also a potential discovery instrument in its own right.

In summary, VLBI provides a unique and powerful complement to conventional SETI techniques. The combination of RFI rejection, precise localisation, and the ability to directly associate signals with astrophysical objects ensures that VLBI with the SKA will be a critical component of any credible technosignature detection and verification strategy.

%\section{Synergies with Other facilities}

\section{Summary}
One of the most enduring questions in astronomy is whether life exists elsewhere in the Universe—and, if so, where it might be found. The SKA is being constructed at a pivotal moment for technosignature research. Both \textit{SKA-Low} and \textit{SKA-Mid} will bridge critical gaps in frequency coverage, sky accessibility, and sensitivity relative to current experiments. Coupled with advances in high-speed computing and the development of algorithms capable of harnessing this power, we now have the means to explore vast datasets with unprecedented depth and precision.

The diverse observing modes available through the \textit{AA*} framework will support a wide range of experimental approaches, enabling studies that push the limits of our understanding of electromagnetic radiation propagation and emission mechanisms. Ultimately, the pursuit of technosignatures with the SKA will not only expand the frontiers of this field but also provide transformative insights into astrophysics and our place in the cosmos.

\bibliographystyle{abbrvnat-maxbibnames4}
\bibliography{chapter} % if your bibtex file is called example.bib

\end{document}